\documentclass[aps,prb,longbibliography,showpacs,floatfix,superscriptaddress,twocolumn]{revtex4-1}

\usepackage{graphicx,color}
\usepackage{amsfonts}
\usepackage[figuresright]{rotating}
\usepackage{amssymb}
\usepackage{amsmath}
\usepackage{mathtools}
\usepackage{psfrag}
\usepackage{subfigure}
\usepackage{multirow}
\usepackage{tabularx}
\usepackage{textcomp}
\usepackage{units}
\usepackage{lipsum}
\usepackage{soul}
\usepackage{titlesec}
\usepackage{times}
\usepackage{hyperref}
\usepackage{enumitem}

\DeclareMathAlphabet\mathbfcal{OMS}{cmsy}{b}{n}

\titlespacing*{\section}
{0pt}{1ex plus .25ex}{1ex plus 1ex}
\titlespacing*{\subsection}
{0pt}{1ex plus .25ex}{1ex plus 1ex}
\titlespacing*{\subsubsection}
{0pt}{1ex plus .25ex}{1ex plus 1ex}

\titleformat{\section}
 {\fontsize{10}{15}\bfseries }
  {\thesection}
  {1em}
  {}

\titleformat{\subsection}
  {\bfseries }
  {\thesubsection}
  {2em}
  {}

\titleformat{\subsubsection}
  {\itshape}
  {\thesubsubsection}
  {2em}
  {}

\def\beq{\begin{eqnarray}}
\def\eeq{\end{eqnarray}}

\newcommand{\bd}[1]{\boldsymbol{ #1 }} % for average
\let\baraccent=\= % rename builtin command \= to \baraccent
\renewcommand{\=}[1]{\stackrel{#1}{=}} % for putting numbers above =
\newcommand{\mc}[1]{\mathcal{ #1}} % mathcal
\makeatletter

\titleclass{\subsubsubsection}{straight}[\subsection]

\newcounter{subsubsubsection}[subsubsection]
\renewcommand\thesubsubsubsection{\thesubsubsection.\arabic{subsubsubsection}}
\titleformat{\subsubsubsection}
   {\normalfont\normalsize\itshape \centering}{\thesubsubsubsection}{1em}{}
\titlespacing*{\subsubsubsection}
{0pt}{1ex plus .3ex}{1ex plus .3ex}
\makeatletter
\renewcommand\paragraph{\@startsection{paragraph}{5}{\z@}%
  {3.25ex \@plus1ex \@minus.2ex}%
  {-1em}%
  {\normalfont\normalsize}}
\renewcommand\subparagraph{\@startsection{subparagraph}{6}{\parindent}%
  {3.25ex \@plus1ex \@minus .2ex}%
  {-1em}%
  {\normalfont\normalsize}}
\def\toclevel@subsubsubsection{4}
\def\toclevel@paragraph{5}
\def\toclevel@paragraph{6}
\def\l@subsubsubsection{\@dottedtocline{4}{7em}{4em}}
\def\l@paragraph{\@dottedtocline{5}{10em}{5em}}
\def\l@subparagraph{\@dottedtocline{6}{14em}{6em}}
\makeatother

\begin{document}
\title{Quantum skyrmion Hall effect}
\author{Ashley M.\ Cook$^*$}
\affiliation{Max Planck Institute for Chemical Physics of Solids, Nöthnitzer Strasse 40, 01187 Dresden, Germany}
\affiliation{Max Planck Institute for the Physics of Complex Systems, Nöthnitzer Strasse 38, 01187 Dresden, Germany}

\begin{abstract}
We consider the problem of magnetic charges in $(2+1)$ dimensions for a torus geometry in real-space, subjected to an inverted Lorentz force due to an external electric field applied normal to the surface of the torus. We compute the Hall conductivity associated with transport of these charges for the case of negligible gapless excitations and global $\mathrm{U}(1)$ charge conservation symmetry, and find it is proportional to an integer-valued topological invariant $\mc{Q}$, corresponding to a magnetic quantum Hall effect (MQHE). We identify a lattice model realizing this physics in the absence of an external electric field. Based on this, we identify a generalization of the MQHE to be quantized transport of magnetic skyrmions, the quantum skyrmion Hall effect (QSkHE), with a $\mathrm{U}(1)$ easy-plane anisotropy of magnetic skyrmions and effective conservation of charge associated with magnetic skyrmions yielding incompressibility, provided a hierarchy of energy scales is respected. As the lattice model may be characterized both by a total Chern number and the topological invariant $\mc{Q}$, we furthermore outline a possible field theory for electric charges, magnetic charges, and correlations between magnetic and electric charges approximated as composite particles, on a two-torus, to handle the scenario of intermediate-strength correlations between electric and magnetic charges modeled as composite particles. We map this problem to a generalized $(4+1)$D theory of the quantum Hall effect for the composite particles.\end{abstract}
\maketitle

    \section{Introduction}

The quantum Hall effect (QHE), in which Hall conductivity quantizes to rational numbers in units of fundamental constants for a two-dimensional electron gas subjected to an external magnetic field applied normal to the plane of the gas, is foundational to understanding of the role of topology in condensed matter. Remarkably, its experimental signatures~\cite{PhysRevLett.45.494, PhysRevLett.48.1559} have been characterized effectively by theoretical developments~\cite{kallin1984excitations, halperin1993theory, halperin1982quantized, halperin1984statistics, wen1991gapless}. In particular, Laughlin's discovery of wavefunctions capturing this physics~\cite{PhysRevLett.50.1395} ultimately led to a microscopically-derived field theory for the QHE~\cite{PhysRevLett.62.82} and a theory of composite fermions for understanding of many fractional quantum Hall states~\cite{PhysRevLett.63.199}. The theory has also very successfully been applied to understanding of multi-component quantum Hall systems ~\cite{PhysRevLett.64.1313, PhysRevB.47.16419} and, more recently, to understanding of the topological insulators and semimetals~\cite{PhysRevB.78.195424, Ryu_2010, schnyder2008}.

Many of these concepts of non-trivial topology of the QHE have also been applied to understanding of topological magnetization textures, known as magnetic skyrmions~\cite{bogdanov1989thermodynamically, yu_real-space_2010, doi:10.1126/science.1166767, schulz_emergent_2012}. While experimental realization and manipulation of magnetic skyrmions has advanced considerably, most study is strongly guided by experiments, which remain focused on overcoming challenges in realization, detection, and manipulation of large, classical, topological magnetization textures~\cite{doi:10.1126/science.1145799, sampaio_nucleation_2013, donnelly_time-resolved_2020, zhang2020skyrmion, maccariello_electrical_2018}. Topological transport signatures have therefore focused primarily on exploiting the classical magnetization textures to generate the quantum anomalous Hall effect via strong Hund's coupling between local moments and itinerant electrons via the topological Hall effect~\cite{PhysRevLett.102.186602, bruno2004topological, PhysRevLett.102.186601}, or transverse motion of classical magnetic skyrmions in response to external fields in the skyrmion Hall effect~\cite{jiang_direct_2017, PhysRevLett.107.136804}, with unquantized transport given the classical nature of the magnetic skyrmions considered.

In this work, we generalize theory of the QHE to identify topological transport signatures of insulating and also itinerant magnets, in which magnetic---rather than electric---charges exhibit quantized transport in response to an external electric---rather than magnetic---field. Such magnetic charges are not relevant to Maxwell's equations in the vacuum, but magnetic monopoles are realized in condensed matter settings as magnetic skyrmions~\cite{castelnovo_magnetic_2008, yu_real-space_2010, doi:10.1126/science.1166767, schulz_emergent_2012}. Rather than classical magnetic skyrmions, we consider possible transport of quantum magnetic skyrmions~\cite{sotnikov2018quantum}. We first approximate them as point magnetic charges, with a global $\mathrm{U}(1)$ magnetic charge conservation symmetry. We consider the problem of transport of such magnetic charges on a two-torus in real-space---rather than the real plane $\mathbb{R}^2$ as  is typically considered in study of magnetic skyrmions---in the presence of an external electric field. This scenario is dual to the QHE~\cite{laughlin1981, niu1985}, and we find quantized transport and incompressibility in the MQHE as a result.

We then identify topological phases of spin, realized in lattice models with spin and orbital degrees of freedom (dof), with the MQHE~\cite{cook2023}. This suggests a possible generalization of the MQHE to quantized transport of magnetic skyrmions, and we identify generalizations of the ingredients in the MQHE to a quantum skyrmion Hall effect (QSkHE), or quantized transport of magnetic skyrmions in response to an external electric field. This requires generalization of the $\mathrm{U}(1)$ charge conservation symmetry important to the incompressibility of the QHE, in particular, to a $\mathrm{U}(1)$ easy-plane anisotropy, corresponding to anisotropic Dzyaloshinskii-Moriya (DM) interactions~\cite{PhysRevB.93.020404, hanada2023babyskyrme}. Based on this, we also generalize the theory of multi-component quantum Hall systems to an effective action containing terms for electric charges and magnetic charges individually coupling to gauge fields, as well as coupling of their correlations, modeled as composite particles, to gauge fields. This theory more robustly handles the scenario of intermediate-strength correlations between electric and magnetic charges, or where electric/magnetic charge conservation individually break down while composite charge conservation is retained.

\section{Results}
\subsection{Magnetic quantum Hall effect}
We consider the problem of magnetic point charges on a two-torus, with $\mathrm{U}(1)$ global charge conservation symmetry, subjected to an external electric field applied normal to the surface of the torus. The external electric field couples to the magnetic charges via an inverted Lorentz force, as Faraday's law is generalized to include a magnetic charge current. Such magnetic charges are realized in condensed matter settings as emergent quasiparticles and topological magnetization textures~\cite{castelnovo_magnetic_2008, yu_real-space_2010, doi:10.1126/science.1166767}. Assuming the $j$\textsuperscript{th} magnetic charge to have mass $\mu_j$ and topological charge $m_j$, the effective Hamiltonian may therefore be written as
\begin{align}
    H &= \sum_{j=1}^N \left[ {1 \over 2 \mu_j} \left( -i \hbar {\partial \over \partial x_j}\right)^2 + {1 \over 2 \mu_j} \left(-i \hbar {\partial \over \partial y_j} - m_j E x_j \right)^2 \right] \\ \nonumber
    &+ \sum_{j=1}^N U(x_j,y_j) + \sum_{\kappa}^N \sum_{\delta}^N V\left( |r_{\kappa} - r_{\delta}| \right),
    \label{eqn1}
\end{align}
where $E$ is the strength of the external electric field applied normal to the surface of the torus, $U$ is an effective background potential, and $V$ is a two-body interaction term. In the following discussion, we neglect interactions governed by $V$ and assume $\mu_j = \mu$ and $m_j = m$ for each $j$. We take boundary conditions to be
\begin{align}
\Gamma_j(L_1, \hat{x})\psi(x_j) &= e^{i \alpha L_1}\psi(x_j), \\
\Gamma_j(L_2, \hat{y})\psi(y_j) &= e^{i \beta L_2}\psi(y_j),
\end{align}
where $j = 1, 2,...,N$ and $\Gamma_j(L_1, \hat{x})$ and $\Gamma_j(L_2, \hat{y})$ are the single-particle electric translation operators in the $x$ and $y$ directions.

\subsection{Quantized transport of magnetic charges}
We start from the Kubo formula expression for Hall conductivity considered by Niu~\emph{et al.}~\cite{niu1985}, but for particles of magnetic charge $m$ in the presence of electric field $\bd{E}$ and background magnetization $n(x,y)$ effectively encoded as a state of $H$,
\begin{equation}
\sigma_m = {i m^2 \over A \hbar} \sum_{n(>0)} {(\nu_1)_{0n}(\nu_2)_{n0}-(\nu_2)_{0n}(\nu_1)_{n0}\over \left( E_0 - E_n \right)^2}.
\label{Hallcond1}
\end{equation}
where $A=L_1 L_2$ is the system size of the torus (of length $L_1$ in the $x$-direction, and $L_2$ in the $y$-direction), $n$ labels excited states, $(\nu_{i})_{0n}$ is the velocity matrix element for the $i$\textsuperscript{th} direction between the ground state $|\phi_0 \rangle$ and excited state $|\phi_n \rangle$, $E_0$ is the energy of the ground state, and $E_n$ is the energy of the $n$\textsuperscript{th} excited state.

We perform a unitary transformation as $\phi_n = exp \left[ -i \alpha(x_1 + \ldots +x_N )\right] exp \left[ -i \beta(y_1 + \ldots +y_N )\right] \psi_n $, rewriting $\sigma_m$ as
\begin{align}
    \sigma_m &= {i m^2 \over A \hbar} \sum_{n,(>0)} \left({\langle  \phi_0| {\partial \tilde{H} \over \partial \alpha}  |  \phi_n\rangle \langle \phi_n | {\partial \tilde{H} \over \partial \beta}   | \phi_0\rangle \over (E_0 - E_n)^2} \right. \\ \nonumber
    &-\left. { \langle  \phi_0|{\partial \tilde{H} \over \partial \beta}   |  \phi_n\rangle \langle \phi_n |{\partial \tilde{H} \over \partial \alpha}   | \phi_0 \rangle \over (E_0 - E_n)^2} \right)
\end{align}
where $\tilde{H}$ is the transformed Hamiltonian.

Using the transformed velocity operators in terms of coordinates $\alpha$ and $\beta$, and then taking $\theta=\alpha L_1$ and $\varphi = \beta L_2$, we simplify the expression to the following,
\begin{equation}
\sigma_m = {i m^2 \over \hbar} \left[ \langle {\partial \phi_0 \over \partial \theta}| {\partial \phi_0 \over \partial \varphi}\rangle- \langle {\partial \phi_0 \over \partial \varphi}| {\partial \phi_0 \over \partial \theta}\rangle \right]
\label{Hallcond2}
\end{equation}

As in Niu~\emph{et al.}~\cite{niu1985}, we assume a finite energy gap separating the ground state from excited states. $\sigma_m$ is computed by averaging over $\theta$ and $\varphi$ as
\begin{align}
\sigma_m &= \bar{\sigma}_m \\ \nonumber
&= {m^2 \over h}\int_0^{2\pi} \int_0^{2\pi} d \theta d \varphi {1 \over 2 \pi i}\left[ \langle {\partial \phi_0 \over \partial \theta}| {\partial \phi_0 \over \partial \varphi}\rangle- \langle {\partial \phi_0 \over \partial \varphi}| {\partial \phi_0 \over \partial \theta}\rangle \right]
\label{Hallcond2avg}
\end{align}

As $\theta$ or $\varphi$ changes by $2 \pi$, the ground state must return to itself (up to overall phase factor), unless the ground state is not uniquely determined by the periodic boundary conditions of the torus. $\bar{\sigma}_m$ is then quantized to integer $\mc{Q}$ times $m^2 / h$, for non-degenerate ground state and finite magnetization gap.

\subsection{Incompressibility}
These arguments may also be applied in the case of open boundary conditions in one direction, and periodic boundary conditions in the second, to revisit Laughlin's gedanken experiment~\cite{laughlin1981} and investigate whether the magnetic charges form an incompressible liquid. We first look for the solution to the following problem of the magnetic point particles for the simplest case of magnetic point particles with fermionic exchange statistics (setting $c=1$):
\begin{equation}
\mc{H} = \sum_j^N \left[{1\over 2 \mu} \left({\hbar \over i} \vec{\nabla}_j - {m } \vec{A}_m(\vec{r}_j) \right)^2 + B m y_j \right],
\end{equation}
where here $\vec{A}_m (\vec{r}) = E y \hat{x}$ is the gauge field for the case of magnetic charges, yielding out-of-plane electric field of strength $E$.

Here, we measure lengths in multiples of $\ell = \sqrt{{\hbar \over \mu \omega_c}} = \sqrt{{\hbar\over m E}}$, where $\omega_c = { m E \over \mu }$. The solution is a Slater determinant of orbitals,
\begin{align}
    \psi_{k,\gamma} = {1 \over \sqrt{2^{\gamma} \gamma !  \sqrt{\pi} L_x}} e^{i k x} e^{(y+y_0 - k)^2/2} \left( {\partial \over \partial y} \right)^{\gamma} e^{-(y + y_0 -k)^2},
\end{align}
where $y_0 = m B l/\hbar \omega_c$. The corresponding energies are
\begin{equation}
E_{k, \gamma} = (\gamma + 1/2)\hbar \omega_c + \hbar  k \left({B \over E} \right) - {\mu  \over 2} \left({B \over E} \right)^2.
\end{equation}
We consider an in-plane magnetic field $B$ is small relative to the gap between these effective Landau levels, and assume a finite energy barrier to gapless excitations (the magnetic charge equivalent of a chemical potential being adjusted to lie within the gap between Landau levels). The number of magnetic charges in the sample is $N = \gamma L_x L_y / 2 \pi l^2$, the magnetic charge density is $\rho_m = {\gamma m \over 2 \pi l^2}$.

The Hall conductivity for transport of the magnetic charges, in terms of magnetic charge current $J_m$, is then
\begin{align}
    \sigma_{m,xy} = {J_m \over B} = {\rho_m  \over E} = {\gamma m^2  \over h}
\end{align}
We therefore find that
\begin{equation}
    J_m = {\gamma m^2 \over \hbar} B.
\end{equation}

Taking the gauge field $\vec{A}_m \rightarrow \vec{A}_m + A_{m,o} \hat{x}$, with $A_{m,o}$ constant, we may further write the current operator as the derivative of the Hamiltonian $\mc{H}$ with respect to the vector potential,
\begin{align}
    {m \over \mu} \sum_j \left[ {\hbar \over i} {\partial \over \partial x}  - {m } A_x(\vec{r}_j) \right] = - {\partial \mc{H} \over \partial A_{m,o}}
\end{align}
If the sample is periodic in one direction as we consider here, $A_{m,o}$ is an electric flux $\phi_E = A_{m,o} L_x$ threaded through the loop.

We may solve the eigen problem of the Hamiltonian for a given $\phi_E$ as $\mc{H}_{\phi_E} = |\Psi_{\phi_E} \rangle = \lambda_{\phi_E} |\Psi_{\phi_E} \rangle$. By Hellman's theorem, we have $\langle \Psi_{\phi_E} | {\partial \mc{H}_{\phi_E} \over \partial \phi_E} |\Psi_{\phi_E} \rangle = {\partial \lambda_{\phi_E} \over \partial \phi_E}$.

Assuming current $J_m$ changes negligibly during the flux insertion, we may write $J_m =  \Delta E / \Delta \phi_E$, with $\Delta \phi_E$ the electric flux quantum $h / m$. Under these conditions, $\mc{H}_{\Delta \phi}$ equals $\mc{H}_0$ up to a gauge transformation. As $\phi_E$ is tuned from $0$ to the electric flux quantum $\Delta \phi_E$, one state (one magnetic charge) per effective Landau level is transferred from the left side of the sample to the right side. Turning on a sufficiently small impurity potential such that gaps remain between the effective Landau levels, adiabatic evolution of $\phi_E$ still transfers exactly one magnetic charge.

\subsection{Counterpart of MQHE on a lattice}
Given that it is possible to realize topological phases characterized by non-trivial Chern numbers without an external magnetic field in lattice-based systems~\cite{haldane1988model}, it is expected that a lattice-based counterpart of the magnetic quantum Hall effect can be realized without external electric field. We provide a minimal toy model for the topological skyrmion phases of matter~\cite{cook2023} as an example, as the topological skyrmion phases of matter are the lattice-based counterparts of the MQHE and a corresponding generalization discussed in the Results, section E. For a system with multiple dofs, these are topological phases associated with subset(s) of these dofs: they have so far specifically been realized in lattice models as topological phases associated with the spin dof in systems also possessing an orbital dof and/or a generalized particle-hole dof. They are characterized by the formation of a magnetic skyrmion in the ground state spin expectation value texture over the Brillouin zone, even when spin is not conserved, due to an additional orbital degree of freedom and non-negligible atomic spin-orbit coupling. We discuss them here as candidates for lattice counterparts of the MQHE without an external electric field, in itinerant magnets.

We first introduce some helpful expressions for three different embeddings of vectors of Pauli matrices into $3 \times 3$ matrix representations, $\boldsymbol{\sigma}_{\alpha} = \langle \sigma_{\alpha,x}, \sigma_{\alpha,y}, \sigma_{\alpha,z}  \rangle$, with components defined as
 \begin{align}
\sigma_{\alpha,x} &= \begin{pmatrix}
0 & \delta_{\alpha 1} & \delta_{\alpha 2} \\
\delta_{\alpha 1} & 0 & \delta_{\alpha 3} \\
\delta_{\alpha 2} & \delta_{\alpha 3} & 0 \\
\end{pmatrix},
\sigma_{\alpha,y} = i\begin{pmatrix}
0 & -\delta_{\alpha 1} & -\delta_{\alpha 2} \\
\delta_{\alpha 1} & 0 & -\delta_{\alpha 3} \\
\delta_{\alpha 2} & \delta_{\alpha 3} & 0 \\
\end{pmatrix} \\ \nonumber
\sigma_{\alpha,z} &= \mathrm{diag}(
\delta_{\alpha 1}+\delta_{\alpha 2}, -\delta_{\alpha 1},-\delta_{\alpha 2}),
\label{sigma1}
 \end{align}
where $\delta_{\alpha \beta} = 1$ if $\alpha = \beta$ and zero otherwise.

 We may then write toy models as Hamiltonians of the form
 \begin{equation}
     \mc{H} = \sum_{\bd{k}}\psi^{\dagger}_{\bd{k}}\mc{H}(\boldsymbol{k})\psi^{}_{\bd{k}},
 \end{equation}
 where $\psi^{}_{\bd{k}} = \left(c_{\bd{k},\uparrow,\alpha}, c_{\bd{k},\downarrow,\beta}, c_{\bd{k},\downarrow,\kappa}   \right)^{\top}$, where $c_{\bd{k},\sigma,\ell}$ annihilates a fermion with spin $\sigma$ in orbital $\ell$. The inspiration for these minimal toy models are tight-binding Hamiltonians describing transition metal oxides such as Sr\textsubscript{2}RuO\textsubscript{4} as studied in past work~\cite{cook2023}, and the t\textsubscript{2g} orbitals of transition metal ions subjected to crystal field splitting could serve as the three-fold orbital degree of freedom considered here. We take the spin to be $1/2$ and effective orbital angular momentum to be $1$. The Bloch Hamiltonian $\mc{H}(\bd{k})$ takes the following form:
 \begin{align}
 \mc{H}(\boldsymbol{k}) = \boldsymbol{d}_1(\boldsymbol{k})\cdot \boldsymbol{\sigma}_1 + \boldsymbol{d}_2(\boldsymbol{k})\cdot \boldsymbol{\sigma}_2.
 \label{latticeHam}
 \end{align}
Even richer topology is realized by adding terms $\mc{H}_{3}(\boldsymbol{k}) = \boldsymbol{d}_3(\boldsymbol{k}) \cdot \boldsymbol{\sigma}_3$.

We compute the spin expectation value using the following spin operators introduced in past work~\cite{cook2023},
  \begin{align}
S_{x} ={1\over 2}\begin{pmatrix}
0 & 1 & 1 \\
1 & 0 & 1 \\
1 & 1 & 0 \\
\end{pmatrix} \hspace{1mm}
S_{y} ={1\over 2} \begin{pmatrix}
0 & -i & -i \\
i & 0 & -i \\
i & i & 0 \\
\end{pmatrix} \hspace{1mm}
S_{z} ={1\over 2} \begin{pmatrix}
2 & 0 & 0 \\
0 & -1 & 0 \\
0 & 0 & -1 \\
\end{pmatrix}
\label{spinops}
 \end{align}

 Rich phase diagrams are realized by choosing each $\bd{d}$-vector to individually characterize a two-band Chern insulator. The system can then at least be characterized in terms of a total Chern number $\mc{C}$ and skyrmion number $\mc{Q}$ as in past work~\cite{cook2023, liu2020}. Here, we consider the example of $\bd{d}_1$ to be a modified QWZ $\bd{d}$-vector~\cite{qi2006_QWZmodel}
 \begin{align}
d_{1,x}(\bd{k}) &= 2\sin (k_y) \\
d_{1,y}(\bd{k}) &= 2\sin (k_x) \\
d_{1,z}(\bd{k}) &= m_1 - 2\cos(k_x) - 2\cos(k_y) ,
 \end{align}
and $\bd{d}_2$ to be the same as $\bd{d}_1$ except for exchange of sine functions and cosine functions (e.g., $\sin{k_x} \rightarrow \cos(k_x)$) and a second, independent mass $m_2$ in place of $m_1$. We also include a term $\lambda \sigma_{3,x}$ in this example.

\begin{figure}[t]
\includegraphics[width=0.5\textwidth]{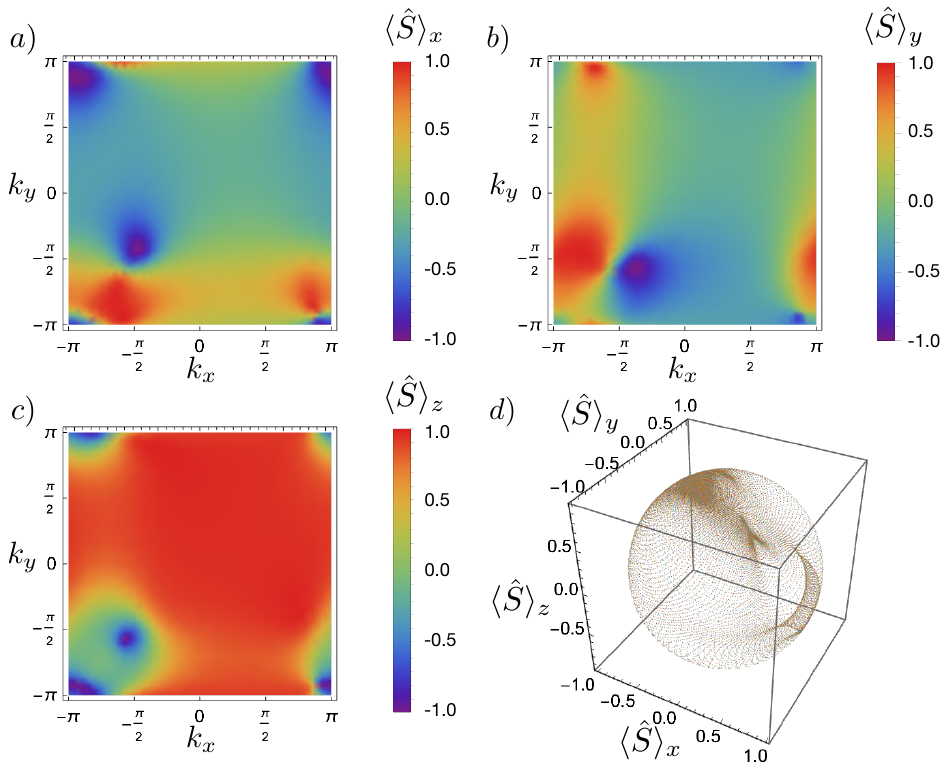}
\caption{Characterization of spin topology in the bulk: winding of $x$, $y$ and $z$ components of normalized ground state spin expectation value texture $\langle \hat{S} \rangle$ over the Brillouin zone for lowest band occupied are shown in a), b), and c), respectively, for $m_1 = m_2 = -1.8$, $\lambda = 0.6$, with topological charge of $\mc{Q}=2$. d) Mapping of normalized ground state spin expectation value texture over the Brillouin zone to the surface of a two-sphere for same coordinates.}
\label{fig1}
\end{figure}

We first compute the topological charge $\mc{Q}$ of the ground state spin expectation value texture over the Brillouin zone, for the lowest band occupied, as
\begin{align}
\mc{Q} = {1 \over 4 \pi} \int d\bd{k} \langle \hat{S} (\bd{k}) \rangle \cdot \left(\partial_{k_x} \langle \hat{S} (\bd{k}) \rangle \times \partial_{k_y} \langle \hat{S} (\bd{k}) \rangle\right),
\end{align}
where $\langle \hat{S} (\bd{k}) \rangle$ is the normalized ground state spin expectation value at momentum $\bd{k}$. We first compute $\langle 1, \bd{k}| S_i |1,\bd{k} \rangle$ for $i \in \{x, y, z \}$ and $S_i$ the spin operators given in Eqn.~\ref{spinops}, with $|1,\bd{k} \rangle$ the lowest-energy eigenstate at momentum $\bd{k}$. We show an example of the spin texture winding in Fig.~\ref{fig1}, corresponding to $\mc{Q}=2$.

\begin{figure}[t]
\includegraphics[width=0.48\textwidth]{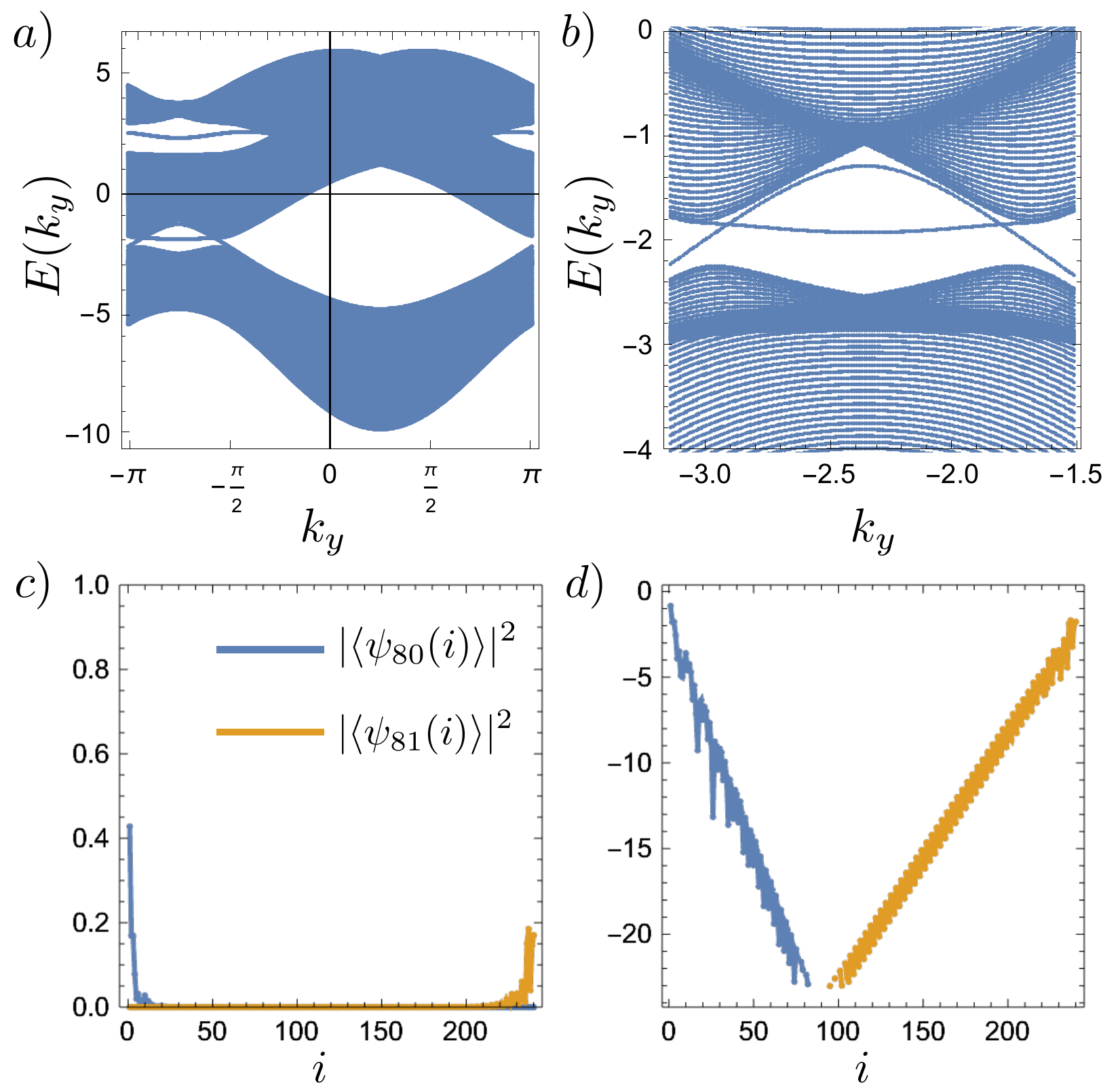}
\caption{a) Full slab energy spectrum for open boundary conditions in $\hat{x}$-direction (system size $L_x=80$) and periodic boundary conditions in the $\hat{y}$ direction. b) slab energy spectrum for same boundary conditions as in a), over smaller interval in slab BZ, highlighting gapless boundary modes. c) Probability density for gapless boundary modes at $k_y = -2.5$ highlighted in b) versus eigenvector index $i$, d) Natural logarithm of probability densities for gapless boundary modes shown in b) and c), illustrating exponential decay.}
\label{fig2}
\end{figure}

For this parameter set, we then open boundary conditions in the $\hat{x}$-direction, and compute the slab energy spectrum, shown in Fig.~\ref{fig2} a). Gapless boundary modes in the slab spectrum are high-lighted in Fig.~\ref{fig2} b). They are not chiral, corresponding to total Chern number of zero, and do not fully traverse the gap, but possess gaplessness as two topologically-protected crossings. The probability density and the natural logarithm of the probability density for each in-gap state is plotted as a function of eigenvector index in Fig.~\ref{fig2} c) and d), respectively, showing exponential localization of the edge states, despite proximity of one of the in-gap states to bulk states for the chosen $k_y = -2.5$.

\begin{figure}[t]
\includegraphics[width=0.48\textwidth]{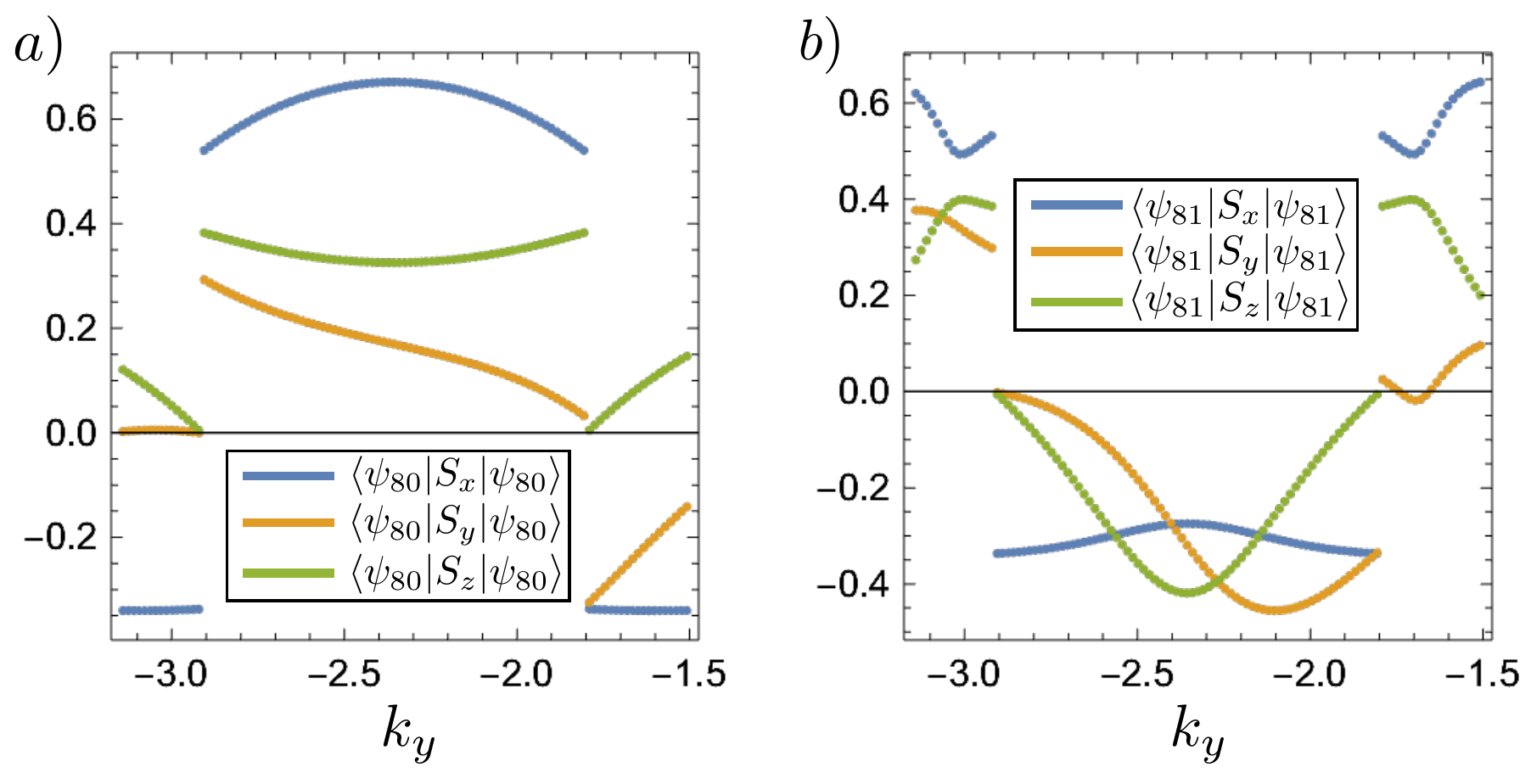}
\caption{Spin expectation value components vs. $k_y$ for gapless boundary modes a) $|\psi_{80}\rangle$  and b) $|\psi_{81}\rangle$, respectively, shown in bulk gap in Fig.~\ref{fig2} b).}
\label{fig3}
\end{figure}

We compute the unnormalized spin expectation value components for these two edge states in Fig.~\ref{fig3} a) and b), respectively. We observe a net pumping of the $\hat{y}$-component of the spin angular momentum over the interval between crossings, corresponding to pumping of spin angular momentum from the bulk valence to bulk conduction bands. At the same time, the structure of the edge states yields no net pumping of charge across the bulk gap, corresponding to total Chern number of zero.

\begin{figure}[t]
\includegraphics[width=0.49\textwidth]{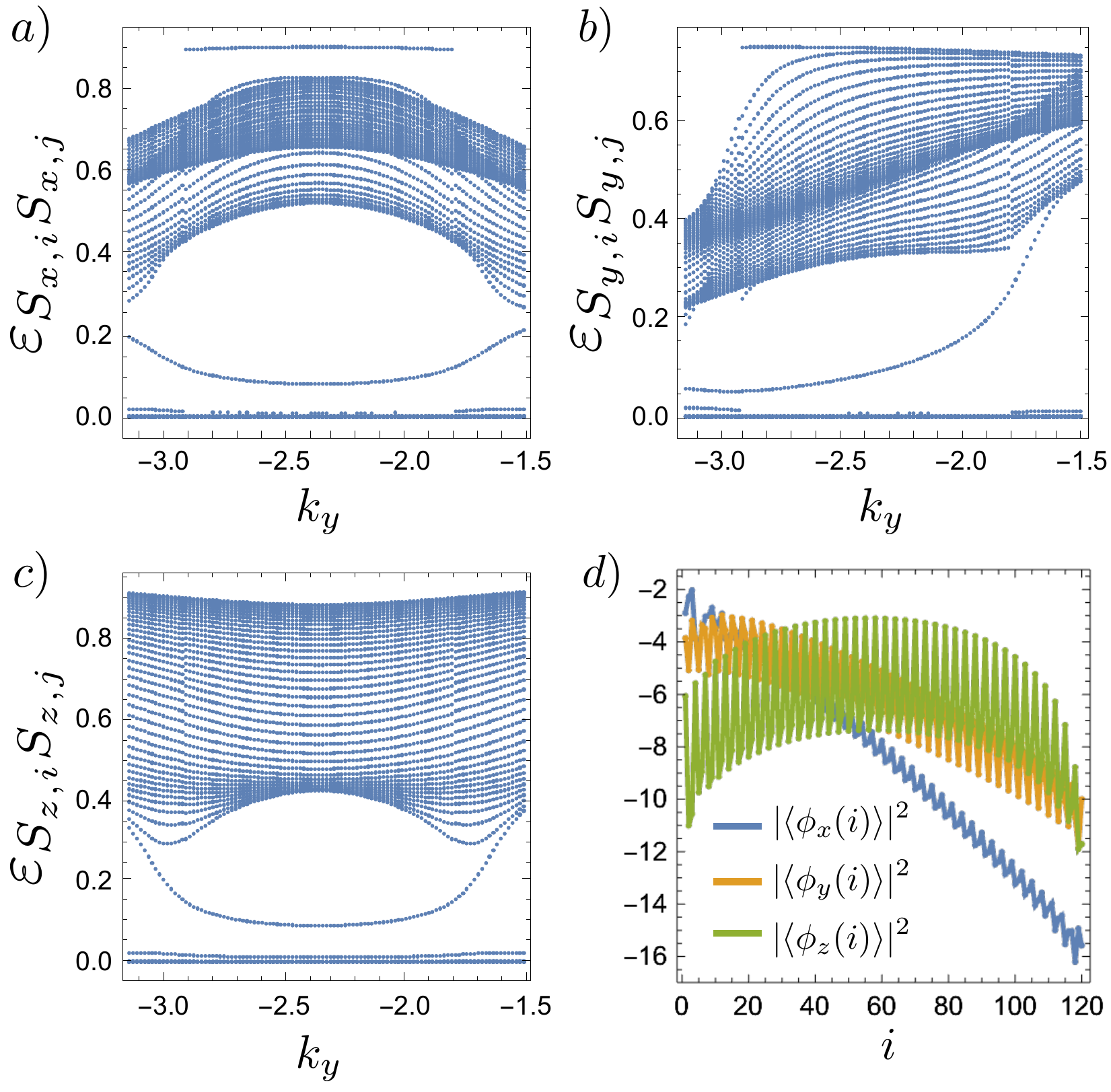}
\caption{Two-point, equal-time spin correlator entanglement spectra for a) $\langle S_{x,i}S_{x,j} \rangle$, b) $\langle S_{y,i}S_{y,j} \rangle$, c) $\langle S_{z,i}S_{z,j} \rangle$, d) Natural logarithm of probability density versus eigenvector index $i$ for highest energy state $|\phi_x \rangle$, $|\phi_y \rangle$, $|\phi_z \rangle$ in each of a), b), c), respectively. Parameter values are $\lambda = 0.6$, $m_1 = m_2 = -1.8$, and $k_y = -2.8$.}
\label{fig4}
\end{figure}

We compute entanglement spectra of two-point, equal-time spin correlators as a generalization of Peschel's method~\cite{IngoPeschel_2003} for this cylinder geometry and band occupancy, tracing out half of the cylinder, shown in Fig.~\ref{fig4}. Over the interval in $k_y$ between the points at which the edge states cross, we observe a state strongly-localized on one edge of the cylinder for the $\hat{x}$- and $\hat{y}$- correlators, but not for the $\hat{z}$-correlator. However, the edge state in the $\hat{y}$-correlator merges with the bulk, corresponding to pumping of the $\hat{y}$-component of spin angular momentum observed in Fig.~\ref{fig3}, while the edge state in the $\hat{x}$- correlator does not merge with the bulk, corresponding to no pumping of the $\hat{x}$-component of spin angular momentum across the cylinder. Neither of these edge states is present for periodic boundary conditions in both the $\hat{x}$- and $\hat{y}$-directions, indicating they are due to bulk-boundary correspondence of non-trivial $\mc{Q}$. We also note that open boundary conditions in the $\hat{y}$-direction and periodic boundary conditions in the $\hat{x}$-direction would instead yield pumping of the $\hat{x}$-component of the spin angular momentum.

\begin{figure}[t]
\includegraphics[width=0.48\textwidth]{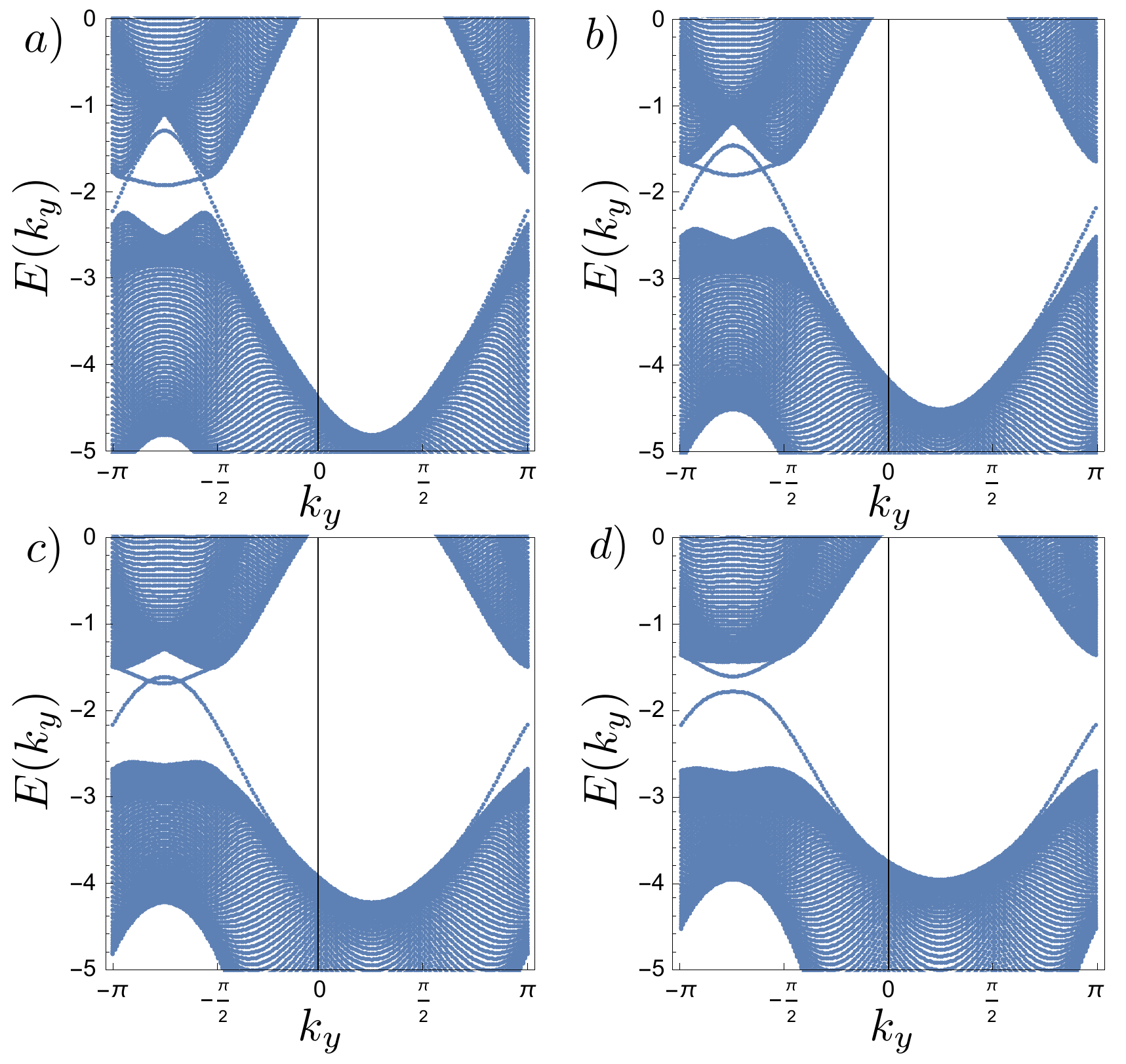}
\caption{Slab energy spectra through type-II topological phase transition (without closing of minimum direct bulk energy gap) from $\mc{Q}=2$ to $\mc{Q}=1$ for $\lambda = 0.6$ and a) $m_1 = m_2 = -1.8$, b) $m_1 = m_2 = -1.6$, c) $m_1 = m_2 = -1.4$, d) $m_1 = m_2 = -1.2$.}
\label{fig5}
\end{figure}

We finally briefly examine a type-II topological phase transition across which $\mc{Q}$ changes from $2$ to $1$ for the chosen parameter set. That is, $\mc{Q}$ changes from one integer value to another as a result of the minimum magnitude of the ground state spin expectation value going to zero, without the closing of the minimum direct bulk energy gap: non-negligible atomic spin-orbit coupling and the orbital dof instead yield total angular momentum that is entirely orbital in character for some $\bd{k}$-value(s) in the Brillouin zone. The transition is shown in Fig.~\ref{fig5}. We observe, for this particular parameter set, that the crossings between the edge states occur over an extended region of phase space, rather than being fine-tuned. The in-gap states eventually separate, however, at which point the edge states in the two-point, equal-time spin correlators are lost.

We note that the spin skyrmion characterized by $\mc{Q}$ is retained upon tracing out the orbital degree of freedom. For the case of non-negligible atomic spin-orbit coupling, the spin degree of freedom is coupled to the orbital degree of freedom, and can therefore be interpreted as an open system upon tracing out the orbital degree of freedom. After performing this partial trace over the orbital degree of freedom, the spin degree of freedom is the only degree of freedom. We may then write an effective Bloch Hamiltonian of the spin subsystem in terms of the ground state spin texture, $\langle \boldsymbol{S}(\boldsymbol{k})\rangle$, as $\mc{H}(\boldsymbol{k}) = -\langle \boldsymbol{S}(\boldsymbol{k})\rangle \cdot \boldsymbol{\sigma}$. As is done in the case of a two-band Chern insulator, we may compute a Hall conductivity $\sigma_{xy}$ in terms of this ground state spin texture $\langle \boldsymbol{S}(\boldsymbol{k})\rangle$. As in the case of the two-band Chern insulator, this Hall conductivity is associated with quantized transport of charge, but is proportional to $\mc{Q}$. $\mc{Q}$ may then be interpreted as an effective Chern number for the open spin subsystem. This Hall conductivity may be distinct from that associated with a total Chern number computed for the full three-band model, and the charge transport associated with Hall conductivity of this effective two-band model is physically distinct from the charge transport of the full three-band Bloch Hamiltonian. That is, non-trivial $\mc{Q}$ is associated with quantized transport and bulk-boundary correspondence of this spin charge of the open spin subsystem.

\begin{figure}[t]
\includegraphics[width=0.5\textwidth]{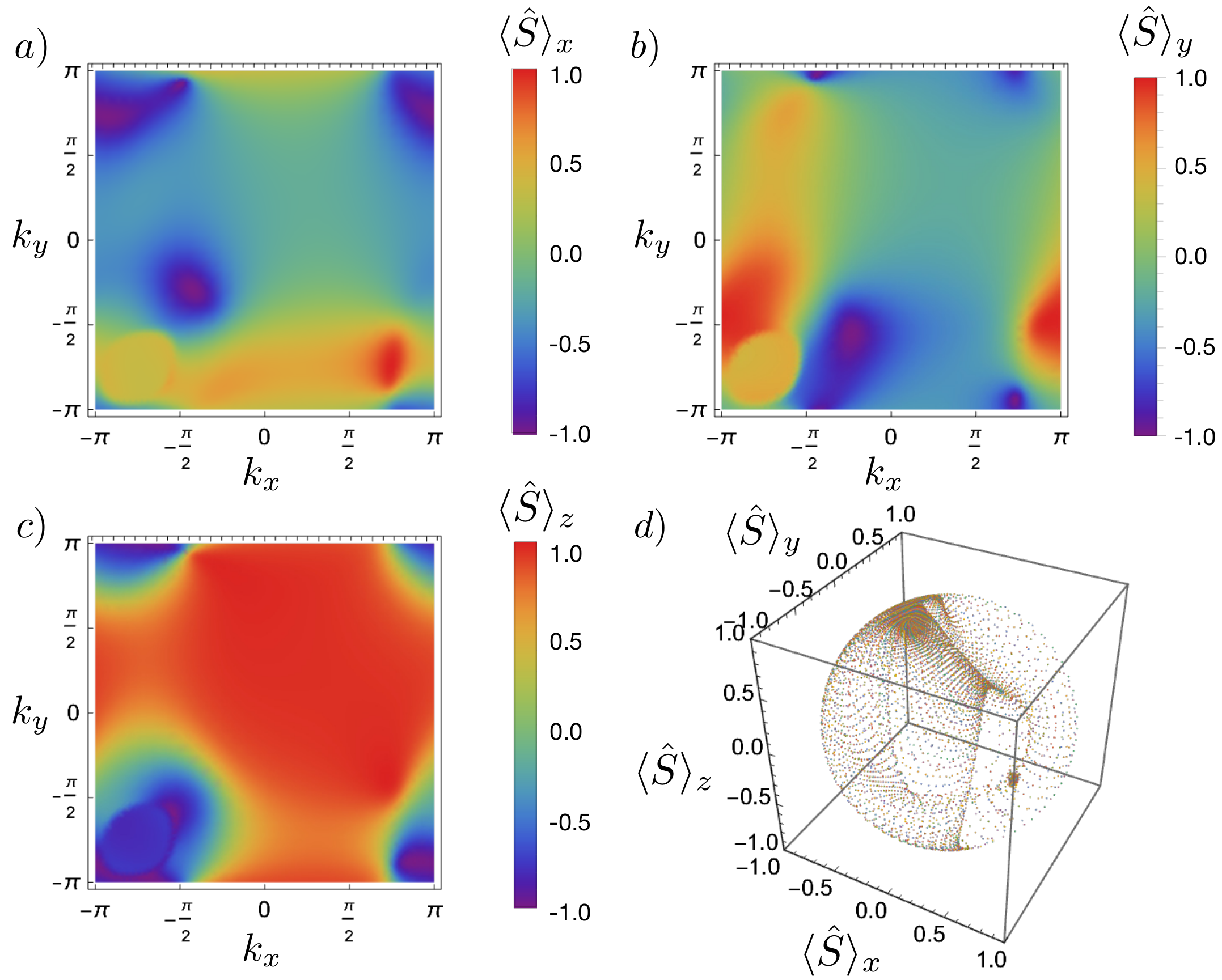}
\caption{Characterization of spin topology for metallic bulk (chemical potential partially intersects second-lowest energy band): winding of $x$, $y$ and $z$ components of normalized ground state spin expectation value texture $\langle \hat{S} \rangle$ over the Brillouin zone for lowest-energy band occupied and second lowest-energy band partially occupied are shown in a), b), and c), respectively, for $m_1 = m_2 = -1.2$, $\lambda = 0.6$, chemical potential $\mu = -1.6$, $k_B T = 0.05$, with topological charge of $\mc{Q}=1$. d) Mapping of normalized ground state spin expectation value texture over the Brillouin zone to the surface of a two-sphere for same coordinates.}
\label{fig6}
\end{figure}

As well, we briefly consider an interesting case of partial filling of the second-lowest energy bulk band. For finite temperature, $\mc{Q}$ can actually be integer-quantized, even for a partially-filled band. We show a representative winding of the ground state spin expectation value components in Fig.~\ref{fig6} a), b), and c), respectively. The normalized ground state spin expectation value texture is also shown mapped to the surface of a two-sphere in Fig.~\ref{fig6} d), demonstrating non-trivial wrapping of the two-sphere. Although the ground state spin expectation value texture for partially-filled bands exhibits singularities in the partial derivatives with respect to $k_x$ or $k_y$, these singularities are removed at finite temperature $k_B T \neq 0$. For spin expectation value finite in magnitude for each $\bd{k}$ in the Brillouin zone, $\mc{Q}$ stabilizes to an integer. That is, $\mc{Q}$ can characterize a topologically stable and non-trivial \textit{metal}. We will explore the consequences of this in greater detail in future work. This is an important demonstration that $\mc{Q}$ is a global topological invariant~\cite{avron1983}.

\subsection{Relevance of MQHE to magnetic skyrmions}
To identify a possible generalization of the MQHE, we define a magnetization texture realizing magnetic skyrmions over a two-torus $T^2$ in real-space of length $L_1$ in the $x$ direction and $L_2$ in the $y$ direction as $M(x, y)$. We first consider $M$ corresponding to $N$ strongly-localized magnetic skyrmions (of radius $r$ much smaller than the minimum of $L_1$ and $L_2$, as well as the spacing between these magnetic skyrmions) against a ferromagnetic background magnetization texture. We assume $N$ is fixed, corresponding to the energy barrier for formation or deletion of these small magnetic skyrmions being much higher than other energy scales of the problem. That is, charge associated with the $N$ small magnetic skyrmions is conserved, provided the energy barrier that must be overcome to create or delete these small magnetic skyrmions---which also includes the energy cost to increase their radius---is much greater than other energy scales in the problem. This scenario might correspond, for instance, to short-range skyrmion density-density interactions dominating over other terms relevant to the problem as minimal ingredients for skyrmion formation, such as ferromagnetic exchange coupling, DM interactions (here we consider anisotropic DM interactions as the out-of-plane external electric field corresponds to a Rashba-type spin-orbit coupling (SOC)~\cite{PhysRevB.93.020404, hanada2023babyskyrme}), and effective Zeeman field term. We note that such anisotropic DM terms yield a $\mathrm{U}(1)$ easy-plane anisotropy of the magnetic skyrmions, which generalizes the $\mathrm{U}(1)$ charge conservation symmetry. That is, the charge conservation symmetry generalizes to conservation of angular momentum. The lattice model provides some evidence of this, exhibiting pumping of spin angular momentum across the bulk gap, without net pumping of charge.

    Although gapless modes, such as phonon modes, may couple to the magnetization texture, we assume the system is at sufficiently low temperature such that these modes are so low in energy that they do not couple to the high-energy internal degrees of freedom of these small magnetic skyrmions.  For sufficiently small magnetic skyrmions, we may furthermore treat them as quantum mechanical, with quantized effective angular momentum associated with their curvature. Such quantum magnetic skyrmions may have already been realized in past work~\cite{sotnikov2018quantum}.

%$+ \int_{T^2} dx dy \Omega_n(x,y)$, and $\Omega_n(x,y)$ is the curvature of a contribution to $M(x,y)$, $n(x,y)$, which serves as an effective background magnetization for the magnetic charges.

    For this magnetization texture possessing $N$ small magnetic skyrmions, we may integrate the curvature associated with the $j$\textsuperscript{th} magnetic skyrmion to yield its topological charge $m_j$. Given a surrounding ferromagnetic magnetization texture and non-negligible exchange coupling, we also assign it an effective mass $\mu_j$: propagation of a small magnetic skyrmion involves deformation of the magnetization texture, and this deformation costs energy due to terms in the Hamiltonian for the magnetization texture, such as the exchange coupling, spin-orbit coupling, and Zeeman field terms. The effective mass in the language of the magnetization texture is the factor by which the velocity of a small magnetic skyrmion is converted to an energy cost.

    We will now create an additional magnetic skyrmion, with radius on the scale of the full system, and therefore comparable in scale to $L_1$ and $L_2$. The energy scale of creating this large skyrmion is set by the couplings in the Hamiltonian for the magnetization texture: the large magnetic skyrmion forming at lower energies than the small magnetic skyrmions requires that high-frequency change in orientation of the magnetization be energetically costly in comparison to low-frequency changes in orientation of the magnetization, in combination with skyrmion density-density interactions favoring large inter-skyrmion spacing for the small magnetic skyrmions. Under these conditions, we form an additional large magnetic skyrmion by applying an external electric field normal to the surface of the torus, of sufficient strength to overcome the energy barrier to formation of the large magnetic skyrmion, which is then also of strength weak compared to the energy scale of creating/deleting the small magnetic skyrmions and the energy scale of changing the radius of a skyrmion from $r$ to the scale of $L_1$ or $L_2$ or vice versa: in simple cases, this external electric field couples to the magnetization texture as an additional anisotropic DM term corresponding to Rashba spin-orbit coupling~\cite{PhysRevB.93.020404}, and can generate an additional large magnetic skyrmion in combination with other terms such as exchange coupling and effective Zeeman field.

    The topological charge of the large magnetic skyrmion, or $\boldsymbol{M}_{BG}(x,y)$, must quantize when the magnetization is everywhere finite in magnitude on the two-torus. Defining a normalized magnetization vector $\hat{\boldsymbol{M}}_{BG}(x,y) = \boldsymbol{M}_{BG}(x,y)/|\boldsymbol{M}_{BG}(x,y)|$, we define mappings from the two-torus $T^2$ to the space of normalized magnetization vectors $S^2$. Such mappings are contractible to mappings $S^2 \rightarrow S^2$, with non-trivial homotopy group $\pi_2 (S^2) = \mathbb{Z}$. We may define a curvature of the magnetization as $\Omega_{M_{BG}}(x,y) =  \hat{\boldsymbol{M}}_{BG}(x,y) \cdot \left(\partial_{x}\hat{\boldsymbol{M}}_{BG}(x,y) \times \partial_{y}\hat{\boldsymbol{M}}_{BG}(x,y)  \right)/4 \pi$, and integrate this curvature over the two-torus to find $\int_{T^2}  \Omega_{M_{BG}}(x,y) dA = \mc{Q}_{BG}$ an integer.

    In the language of the MQHE, magnetic charges experience a local electric flux resulting from coupling to the external electric field. In the language of the magnetization texture, the local electric flux after integrating out the internal degrees of freedom of the small magnetic skyrmions is $\Omega_{M_{BG}}(x,y)$. These two curvatures are therefore equal at a given point on the two-torus, despite being computed from different physical quantities.

    Although we can neglect coupling of gapless excitations to the small magnetic skyrmions by considering the case of high-energy internal dofs of the small magnetic skyrmions, and the system at sufficiently low temperatures such that gapless excitations do not couple to these high-energy internal dofs, there remains the possibility that gapless excitations could couple to the large magnetic skyrmion. Considering the system at sufficiently low temperature such that these gapless excitations cannot change the topological charge of this large magnetic skyrmion, $\mc{Q}_{BG}$, the remaining potential concern is the role played by local perturbations of the system due to gapless excitations. This question has already been explored in the context of quasi-topological phases by Bonderson and Nayak~\cite{bonderson2013}: the large magnetic skyrmion is in a ground state protected by finite energy barrier, $\Delta$, to change in its topological charge. Gapless excitations could yield a set of pseudo-groundstates $|a \rangle$, $a=1,2,...,N_{\Sigma}$, in the language of Bonderson and Nayak, with energy $E_{a}$ satisfying the following relation:
    \begin{align}
        \delta E_0 \equiv \mathrm{max}|(E_a - E_{a'})| = \mathcal{O}(e^{-L/\varepsilon}),
    \end{align}
    where $L$ is the system size and $\varepsilon$ is the correlation length of the system. Local perturbations at energies below $\Delta$ act within this manifold of pseudo-groundstates: the system can therefore acquire an overall phase at most and quantum information encoded in this pseudo-ground state subspace is
“topologically protected” at zero temperature in the thermodynamic limit. At non-zero temperature $T$, the error rate $\Gamma$ is
exponentially suppressed by the gap, $\Gamma \sim e^{-\Delta/T}$.

    For our purposes, then, gapless excitations corresponding to changes in local curvature of the magnetization texture, which preserve the topological charge of the large magnetic skyrmion and do not couple to the high-energy internal degrees of freedom of the small magnetic skyrmions, are permitted. In the language of the magnetic charges, this corresponds to local deformation of the electric flux while remaining within a given Landau level.

    Under these conditions, we  identify quantized transport of small magnetic skyrmions with quantized transport of magnetic charges. This quantum skyrmion Hall effect (QSkHE) is therefore a generalization of the MQHE, in that the magnetic charges of the MQHE are generalized to small magnetic skyrmions, the $\mathrm{U}(1)$ charge conservation symmetry of the MQHE is generalized to a $\mathrm{U}(1)$ easy-plane anisotropy and charge conservation symmetry reliant on maintenance of a hierarchy of energy scales, and the background electric flux is generalized to a large background magnetic skyrmion. A lattice counterpart of the QSkHE is the topological phase realized by the lattice model Eqn.~\ref{latticeHam} for $\mc{Q} \neq 0$, which corresponds to the lattice counterpart of the MQHE after partial trace over orbital dof. We also note that the flat-band limit assumption, first stated in its modern form in Avron~\emph{et al.}~\cite{avron1983}, therefore derives from an earlier assumption of point charges in the theory of the QHE~\cite{laughlin1981}, which can be relaxed to generalize the QHE as the QSkHE.

\subsection{Role of external field in the quantum skyrmion Hall effect}
For applied field strength comparable to the energy scale of the large background magnetic skyrmion, changing the external field strength corresponds to flux insertion and pumping of charges to the boundary due to incompressibility. For applied field strength comparable to the energy scale of creating/deleting small magnetic skyrmions, changing the external field strength corresponds to tuning the chemical potential of the magnetic charges, corresponding to proliferation of small magnetic skyrmions due to incompressibility.

\subsection{Two-dimensional gas of both electric and magnetic charges}
We now explore the role of the QSkHE in the context of the quantum Hall problem in $(2+1)$ dimensions, generalized to the case of a 2D gas of electric---but also magnetic---charges. That is, what is the potential role of the QSkHE in the response and topology of an itinerant magnetic system confined to a 2D plane? This investigation is also motivated by past study of topological skyrmion phases in lattice systems, where the Chern insulator phase coexists with topological phases characterized by spin skyrmion number $\mc{Q}$~\cite{cook2023, liu2020}, as well as past work on multiplicative topological phases, which exhibit non-trivial topology of single particle systems and also of the correlations between single particle systems~\cite{cook_multiplicative_2022}.

First, we note that the MQHE is dual to the QHE: exchanging magnetic charges for electric charges, and the applied electric field for a magnetic field in the MQHE, we map to the QHE. We therefore expect the field theory of the itinerant magnet, for the case of both electric field and magnetic field applied normal to the plane of the system, to be self-dual: exchanging magnetic charge for electric charge and external electric field for external magnetic field should yield the same field theory.

Such self-duality is a property of the $(4+1)$ dimensional quantum Hall effect, in that there is a symmetry in the field theory between how particles respond to external magnetic versus electric fields~\cite{qi2008}. We show that the $(4+1)$D QHE can be mapped to the problem of a 2D system of electric and magnetic charges. To do so, we reinterpret the particles coupled to external fields in the $(4+1)$D field theory for the QHE, $q_c$, as composite particles, composed of a particle with electric charge, $q_e$, and a particle with magnetic charge, $q_m$, as $q_c = q_e +q_m$.

By reinterpreting the axes $x$, $y$, $z$, $\omega$ of four-dimensional space as $x$, $y$, $x'$, $y'$, where $x$ and $y$ are the coordinates of the electric charges, and $x'$ and $y'$ are the coordinates of the magnetic charges, we map the $(4+1)$D theory of composite particles to a $(2+2+1)$D theory of electric and magnetic charges of a two-dimensional gas. For consistency in realizing an $\bd{E} \cdot \bd{B}$ coupling for this interpretation of particles in the $(4+1)$D theory as composite, we introduce an operator $f^{\dagger}_{c,\bd{r},\bd{r'}} = f^{\dagger}_{e,\bd{r}}f^{\dagger}_{m,\bd{r}'}$ to create composite particles at position $\bd{r}  = (x,y)$, $\bd{r}' = (x', y')$, where $f^{\dagger}_{e,\bd{r}}$ creates a particle with electric charge at position $\bd{r}$, and $f^{\dagger}_{m,\bd{r}'}$ creates a particle with magnetic charge at position $\bd{r}'$. For electric and magnetic charges each fermionic, the composite particle is then a composite boson~\cite{combescot_15the_2015}.

We consider external magnetic and electric field each applied normal to the plane of the gas, rather than just a magnetic field or electric field. $\bd{E}$ and $\bd{B}$ are therefore parallel. In this case, the $(4+1)$D theory includes the terms of the $(2+1)$D theory of the QHE and MQHE: $\bd{E}$ exerts a force on electric charge, but only in the $\hat{z}$ direction, so this coupling is neglected and only $\bd{B}$ coupling to electric charge is non-trivial. Similarly, $\bd{B}$ can exert a force on the magnetic charges, but only in the $\hat{z}$-direction, such that it is trivial and only the $\bd{E}$ coupling to the magnetic charges is non-trivial. However, with $\bd{E}$ and $\bd{B}$ parallel, the magnetoelectric term of the $(4+1)$D theory of composite particles, corresponding to a topologically non-trivial magnetoelectric response of the system expressed in terms of the second Chern form~\cite{qi2008}, is relevant but unnecessary to fully characterize the topology of the system in simple cases.

We previously considered a $\mathrm{U}(1)$ globally-symmetric MQHE, and its generalization to the QSkHE with $\mathrm{U}(1)$ easy-plane anisotropy due to anisotropic DM interaction: a charge conservation symmetry effectively holds in this case, provided a certain hierarchy of energy scales is maintained, in which the energetic cost of creation/deletion or increasing the radius of a small magnetic skyrmion is significantly greater than other operations. We now consider the scenario where correlations between electric and magnetic charges are intermediate in strength, such that electric and magnetic charges are not necessarily at the same positions, and these correlated states are not necessarily tensor products of states of the electric charge subsystem and states of the magnetic charge subsystem. Additionally, we consider the scenarios in which electric charge conservation and magnetic charge conservation each hold in some parameter regimes, but are potentially lost while composite charge conservation is potentially retained in some parameter regimes.

In these regimes, a third contribution to the action, associated with the correlations between electric and magnetic charges modeled as composite particles, more fully captures the quantum Hall physics of this multi-component system. Such higher-dimensional topology of correlations has been realized previously as multiplicative topological phases~\cite{cook_multiplicative_2022}, in particular as the perpendicular multiplicative Kitaev chain~\cite{pal2023multiplicative}, although these are simple cases where the topology of the correlations derives from the single particle topology of each parent phase.

We therefore consider an effective action $S_{\mathrm{eff}}$ for the $(2+2+1)$D theory in terms of three gauge fields, one for the coupling of external fields to electric charges (capturing the fact that only the $\bd{B}$ field couples in this case), $A_{e,i}$, one for coupling of the external fields to the magnetic charges (capturing the fact that only the $\bd{E}$ field couples to these particles, but \textit{as $\bd{B}$ couples to the electric charges}), $A_{m,i}$, and a gauge field $A_{c,i}$, capturing the coupling of fields to the composite particles, $S_{\mathrm{eff}} = S_{e,\mathrm{eff}} + S_{m,\mathrm{eff}} + S_{c,\mathrm{eff}}$,
where
\begin{align}
S_{e,\mathrm{eff}} &= {C \over 4 \pi} \int  d^2x_e dt \varepsilon^{\mu \nu \rho} A_{e,\mu} \partial_{\nu} A_{e, \rho}\\
S_{m,\mathrm{eff}} &=  {\mc{Q} \over 4 \pi} \int  d^2x_m  dt
 \varepsilon^{\mu \nu \rho} A_{m,\mu} \partial_{\nu} A_{m, \rho} \\
S_{c,\mathrm{eff}} & = {C_2 \over 24 \pi^2} \int d^4x dt \varepsilon^{\mu \nu \rho \sigma \tau} A_{c,\mu} \partial_{\nu} A_{c,\rho}  \partial_{\sigma} A_{c,\tau}
\end{align}
In the $(4+1)$D QHE, one can pump $C_2$ charges by applying magnetic fields corresponding to gauge field $\bd{A} = \left(By, 0, B\omega, 0 \right)$~\cite{qi2008}. In the $(2+2+1)$D theory, this corresponds to $\bd{A}_c = \left(By, 0, E y', 0 \right)$, which will correspond to pumping of $C_2$ electric and magnetic charges, corresponding to pumping of $C_2$ composite particles as required.

\section{Discussion and Conclusion}
We study the problem of magnetic charges in $(2+1)$ dimensions subjected to an external electric field, which couples to the magnetic charges as an inverted Lorentz force. Considering this problem on a two-torus with $\mathrm{U}(1)$ global symmetry corresponding to conservation of this magnetic charge, we find the Hall conductivity governing transport of the magnetic charges quantizes in the case of negligible gapless excitations, being proportional to an integer-valued topological invariant $\mc{Q}$, to realize a \textit{magnetic quantum Hall effect} (MQHE).

We present results on a counterpart lattice problem, realizing physics of the MQHE in the absence of an external electric field, characterized by a three-band Bloch Hamiltonian. We characterize its topology in terms of a total Chern number $\mc{C}$ and a winding number $\mc{Q}$ for the ground state spin expectation value as in past work on topological skyrmion phases~\cite{cook2023}, which we identify as the lattice counterparts of the MQHE. We mean this in the same sense that the Chern insulator is the lattice counterpart of the QHE realized in the absence of an external magnetic field~\cite{haldane1988model}, and note that the topological skyrmion phases discussed here can be understood as generalizations of Chern insulators. We consider a topological phase characterized by total Chern number $\mc{C}=0$ and skyrmion number $\mc{Q}=2$, demonstrating the quantized topological charge of the ground state spin expectation value texture over the bulk Brillouin zone, and a bulk-boundary correspondence in which non-trivial $\mc{Q}$ yields gapless modes localized on the system boundary. Here, gaplessness is ensured by the crossing(s) of edge states in the slab spectrum that permit pumping of spin angular momentum without pumping of net charge.

We develop a generalization of Peschel's method~\cite{IngoPeschel_2003} for computing entanglement spectra, to compute entanglement spectra for two-point, equal-time spin correlators of the lattice model, in which we observe a bulk-boundary correspondence when the gapless edge states are present in the slab energy spectrum: edge states in the entanglement spectra correspond to pumping of the $\hat{y}$-component of spin angular momentum for sample edges oriented in the $\hat{y}$-direction.

We also show an example of a type-II topological phase transition~\cite{cook2023} over which the gapless boundary modes are lost without the closing of the minimum direct bulk energy gap, which may be helpful in efforts to experimentally confirm these topological phases of matter. As well, we demonstrate quantized winding of the ground state spin expectation value in the bulk Brillouin zone for the case of the lattice model describing a bulk \textit{metal} at small but finite temperature. This illustrates the fact that the skyrmion number is a global topological invariant~\cite{avron1983}, as well as potentially a topological invariant for characterizing topological $\textit{metals}$ in addition to semimetals and insulators, which we will explore in greater detail in future work.

As the lattice model realizes a counterpart of the MQHE \textit{after} tracing out an orbital degree of freedom, as well as a generalization of the MQHE \textit{prior} to performing a partial trace, we also identify a generalization of the MQHE without a lattice, the quantum skyrmion Hall effect (QSkHE).  The magnetic point charge generalizes to a small magnetic skyrmion, and the $\mathrm{U}(1)$ charge conservation generalizes to a $\mathrm{U}(1)$ easy-plane anisotropy corresponding to anisotropic DM interactions and charge conservation symmetry maintained by a hierarchy of energy scales, in which the energetic costs of creating/deleting small magnetic skyrmions or increasing their radius are far greater than energetic costs of other operations. At sufficiently low temperatures, gapless excitations do not couple to the high-energy internal dofs of the small magnetic skyrmions, but can yield local perturbations of the large magnetic skyrmion which do not interfere with incompressibility. The lattice model counterpart of the QSkHE and MQHE for a spin $1/2$ dof in a system also possessing an orbital dof and non-negligible atomic SOC furthermore suggests this physics persists in non-equilibrium settings, as the spin dof may be interpreted as an open system, coupling to an orbital dof environment via non-negligible atomic SOC. We also note that the generalization from a $\mathrm{U}(1)$ charge conservation symmetry to the $U(1)$ easy-plane anisotropy corresponds to generalization of pumping of charge to pumping of angular momentum, which is observed in the full lattice model.

We then consider a field theory for an itinerant magnet in $(2+1)$ dimensions inspired by both the topological skyrmion phases~\cite{cook2023} and the multiplicative topological phases~\cite{cook_multiplicative_2022}, which are topological phases of correlations between single-particle systems. This field theory includes terms in the effective action for the QHE of electric charges and the MQHE of magnetic charges relevant in cases where $\mathrm{U}(1)$ charge conservation holds for each type of charge individually, as well as terms of a $(4+1)$D QHE of composite particles consisting of electric and magnetic charge pairs, effectively capturing topology of their correlations relevant in cases of intermediate-strength correlations between electric and magnetic charges, and where the system is not well-described as a direct product of a system of electric charges and a system of magnetic charges.

In the case of the itinerant magnet, it is not clear whether a more comprehensive characterization of topology beyond the three topological invariants considered here is required, and this will be explored in future work. Importantly, the assumption of point-like charges could, more generally, be relaxed, with a more general field theory expressed in terms of underlying polarization and magnetization fields. Related to this, we note that the QSkHE can potentially be realized in vector fields of myriad observables, including the polarization, if these vector fields can satisfy the requirements outlined here.

Past work on the quantum Hall ferromagnet has already shown magnetic skyrmions form in the context of the quantum Hall effect~\cite{PhysRevB.47.16419,girvin_multi}. Our results are likely relevant to this problem. We may also extend these results to a fractional quantum skyrmion Hall effect (FQSkHE), similarly to the treatment of the fractional quantum Hall effect in Niu~\emph{et al.}~\cite{niu1985}, by adjusting system size $A$ relative to magnetic particle number $N$ to realize overall partial filling.

\textit{Acknowledgements}---We would like to thank C.~Nayak, J.~E.~Moore, X.~L.~Qi, S.~Raghu, E.~Altman and S.~A.~Kivelson for helpful discussions. We would also like to thank R.~Moessner for many helpful discussions and careful reading of the manuscript.

\bibliography{main.bib}

\clearpage

%%%%%%%%%% Merge with supplemental materials %%%%%%%%%%
%\pagebreak
%\widetext
%%%%%%%%%% Prefix a "S" to all equations, figures, tables and reset the counter %%%%%%%%%%
%\setcounter{equation}{0}
%\setcounter{figure}{0}
%\setcounter{table}{0}
%\setcounter{page}{1}
%\makeatletter
%\renewcommand{\theequation}{S\arabic{equation}}
%\renewcommand{\thefigure}{S\arabic{figure}}
%\renewcommand{\bibnumfmt}[1]{[S#1]}
%\renewcommand{\citenumfont}[1]{S#1}
%%%%%%%%%% Prefix a "S" to all equations, figures, tables and reset the counter %%%%%%%%%%

\end{document}